\documentclass{emulateapj}

\usepackage{apjfonts}
\usepackage{amsmath}
\usepackage{amssymb}
\usepackage{natbib}
\usepackage{xspace}
\usepackage{graphicx}
\usepackage{rotating} 
\usepackage{subfigure}
\usepackage{float}

\newcommand{\err}[2]{\ensuremath{^{+#1}_{-#2}}\xspace}
\newcommand{\e}[1]{\ensuremath{^{#1}}\xspace}
\newcommand{\ten}[1]{\ensuremath{\times 10^{#1}}\xspace}

\newcommand{\Msun}{\ensuremath{M_\odot}\xspace}

\newcommand{\nh}{$N_{\text{H}}$\xspace}

\newcommand{\pexrav}{\textsc{pexrav}\xspace}
\newcommand{\pexmon}{\textsc{pexmon}\xspace}
\newcommand{\xspec}{\textsc{xspec}\xspace}
\newcommand{\mytorus}{MYT\textsc{orus}\xspace}
\newcommand{\etal}{et al.\xspace}
\newcommand{\ka}{K$\alpha$\xspace}

\newcommand{\chidof}{$\chi^{2}/$dof\xspace}

\newcommand{\xte}{\textsl{RXTE}\xspace}

\newcommand{\sax}{\textsl{BeppoSAX}\xspace}
\newcommand{\suzaku}{\textsl{Suzaku}\xspace}

\newcommand{\bat}{\textsl{Swift}-BAT\xspace}
\newcommand{\xmm}{\textsl{XMM-Newton}\xspace}

\newcommand{\fluxunits}{erg\,cm$^{-2}$\,s$^{-1}$\xspace}
\newcommand{\feunits}{photons\,cm$^{-2}$\,s$^{-1}$\xspace}

\newcommand{\colunits}{cm$^{-2}$\xspace}

\shorttitle{NGC 3660}
\shortauthors{Rivers \etal }

\begin{document}

\title{\textsl{Suzaku} Confirms NGC~3660 is an Unabsorbed Seyfert 2} 
\author{E.~Rivers\altaffilmark{1}, M.~Brightman\altaffilmark{1}, S.~Bianchi\altaffilmark{2}, G.~Matt\altaffilmark{2}, K.~Nandra\altaffilmark{3}, Y.~Ueda\altaffilmark{4}}

\altaffiltext{1}{Cahill Center for Astronomy and Astrophysics, California Institute of Technology, Pasadena, CA 91125, USA} 
\altaffiltext{2}{Dipartimento di Matematica e Fisica, Universit\`a degli Studi Roma Tre, via della Vasca Navale 84, 00146 Roma, Italy}
\altaffiltext{3}{Max Planck Institute for Extraterrestrial Physics, Giessenbachstrasse, 85741 Garching, Germany}
\altaffiltext{4}{Department of Astronomy, Kyoto University, Japan}

\email{Contact: erivers@caltech.edu}

\begin{abstract}

An enigmatic group of objects, unabsorbed Seyfert 2s may have intrinsically weak broad line regions, 
obscuration in the line of sight to the BLR but not to the X-ray corona, or so much obscuration that the X-ray continuum 
is completely suppressed and the observed spectrum is actually scattered into the line of sight from nearby material.
NGC 3660 has been shown to have weak broad optical/near infrared lines, no obscuration in the soft X-ray band, and no indication of ``changing look'' behavior. 
The only previous hard X-ray detection of this source by \textsl{Beppo-SAX} seemed to indicate that the source might harbor a heavily obscured nucleus.  
However, our analysis of a long-look \textsl{Suzaku} observation of this source shows that this is not the case,
and that this source has a typical power law X-ray continuum with normal reflection and no obscuration.  
We conclude that NGC 3660 is confirmed to have no unidentified obscuration and that the anomolously high \textsl{Beppo-SAX} measurement must 
be due to source confusion or similar, being inconsistent with our \textsl{Suzaku} measurements as well as non-detections from \\textsl{Swift-BAT} and \textsl{RXTE}.

\end{abstract}


\section{Introduction}

The standard unification scheme of active galactic nuclei (AGN, e.g., Antonucci 1993) explains the observational differences between 
type 1 and type 2 Seyfert galaxies as being due to the orientation of the nucleus relative to the observer. 
According to this scheme, in Seyfert 1s we see the central engine, an accretion disk surrounding a supermassive black hole, directly. 
This allows full view of the accretion disk, X-ray corona, and broad line region (BLR) where the broad optical lines that characterize type 1 Seyferts are produced.
Meanwhile in Seyfert 2s, the central engine is obscured from sight by an optically-thick torus structure composed of gas and dust 
and only narrow optical emission lines, produced at much larger distances, are observed. 

Support for this scheme includes the detection of broad optical emission lines in the polarized spectra of Seyfert 2s (Antonucci \& Miller 1985) 
and strong X-ray absorption (Awaki \etal 1991). However, it has become clear that orientation effects cannot alone explain the full range of AGN observations. 
For example, a luminosity modification is required to explain why the fraction of obscured sources decreases with luminosity (Ueda \etal 2003). 
Furthermore, accretion rate is likely an important parameter, as the detection rate of the polarized broad lines appears to decrease with accretion rate 
(Nicastro \etal 2003; Marinucci \etal 2012)

An enigmatic group of objects, unabsorbed Seyfert 2s, poses a particular challenge for the unification scheme (Pappa \etal 2001). 
These Seyferts have no detectable optical broad lines, indicating that the central engine is obscured, however absorption is not 
measured in their X-ray spectra, indicating an unobscured sight line to the central engine. The most intriguing and controversial explanation 
is that these sources intrinsically lack the BLR, and hence are known as ``true'' Seyfert 2. Theoretical modeling suggests that the 
BLR may not be present at low luminosities/accretion rates (Nicastro, 2000; Elitzur \& Shlosman, 2006; Trump \etal 2011). 
It is therefore essential to obtain a sizable number of confirmed true Seyfert 2s in order to constrain AGN accretion and unification models. 

Alternative explanations to the true Seyfert 2 scenario include unidentified obscuration, whereby the X-ray spectrum may be contaminated by 
emission from the host galaxy, or strong scattered emission from the AGN, neither of which are expected to be strongly absorbed (e.g. NGC 4501, Brightman \& Nandra 2008). 
Furthermore, a state change is possible, where the source transitions from an obscured state to an unobscured state, as material passes across the line of sight. 
If the X-ray and optical observations are not simultaneous, this could lead to the mismatch (e.g. NGC 1365, Risaliti  \etal 2005). Also possible is an anomalously 
high dust-to-gas ratio, which causes severe reddening of the BLR by dust, while the X-rays are not attenuated by gas (e.g. IRASF01475-0740, Huang \etal 2011).

NGC~3660 was presented as a true Seyfert 2 candidate in the Brightman \& Nandra (2008) sample.
This source has very weak measured broad lines and no evidence of polarized broad optical lines (Tran 2001; Shi \etal 2010; Tran \etal 2011). 
From an \textsl{ASCA} observation, this source displayed rapid X-ray variability on kilosecond scales, 
strong evidence that the source is viewed directly. Subsequently, Bianchi \etal (2012) presented a simultaneous optical/\xmm observation. 
The \xmm spectrum confirmed the lack of X-ray absorption and rapid X-ray variability, while the optical spectrum simultaneously revealed very weak broad lines, 
thus ruling out a state change. Near infrared spectroscopy also failed to show strong broad lines, making a high dust to gas ratio unlikely. 
However, a hard X-ray ($>$10 keV) observation from \sax revealed a 20--100 keV flux of 2.67\ten{-11} \fluxunits, 
well in excess of the 2--10 keV emission extrapolated into the hard band. 
This is suggestive of the possibility that NGC 3660 might in fact be harboring a far more powerful, Compton-thick obscured AGN (Dadina  \etal 2007), 
though this would leave the observed variability unexplained.

In this paper we present analysis of a long-look \suzaku observation of NGC 3660 ($z$=0.012) in the 0.5--35 keV X-ray band.
This broad X-ray band allows us to test whether this galaxy does indeed harbor a heavily obscured AGN.  
This paper is organized as follows: data reduction is presented in Section 2, spectral analysis is described in 
Section 3, and results are presented and discussed in Section 4.


\begin{figure}
      \includegraphics[width=0.47\textwidth]{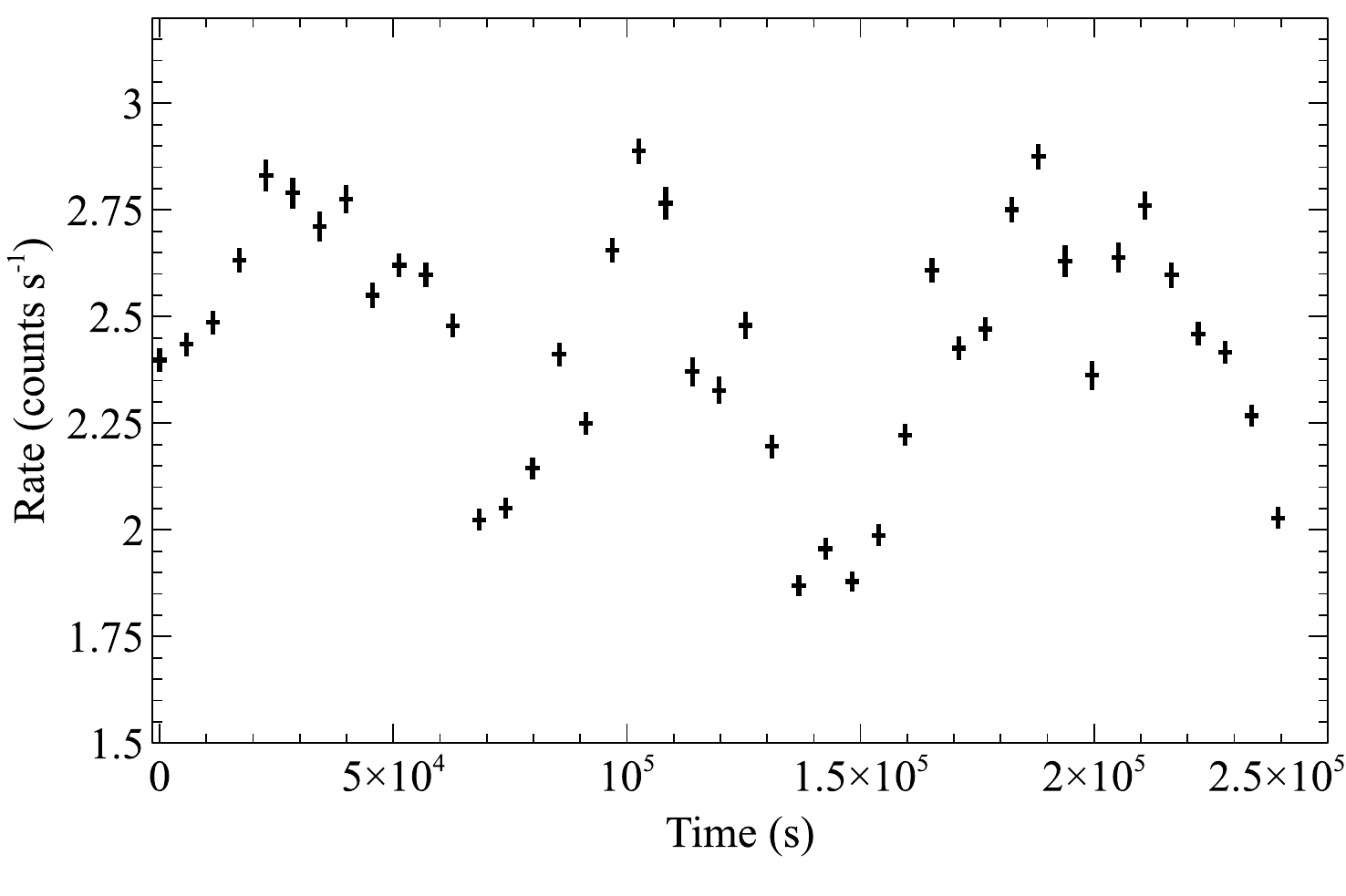}
 \caption{Total XIS count rate in the 0.5--10 keV band binned to 5 ks with an average rate of 2.4 counts s\e{-1}.  The high level of short timescale variability seen here is very similar to that displayed in previous observations (e.g., Bianchi \etal 2012).}
  \label{figlc}
\end{figure}


\section{Data Reduction}\label{sec:analysis}

\textsl{Suzaku} has two pointed instruments, the X-ray Imaging Spectrometer (XIS; Koyama \etal \ 2007) and the Hard X-ray Detector (HXD; Takahashi \etal \ 2007).
Data were taken 2013 Nov 28 -- Dec 01 (OBSID 708043010).
They were processed with version 2.8.20.35 of the \textsl{Suzaku} pipeline and the recommended screening criteria were applied 
(see the \textsl{Suzaku} Data Reduction Guide\footnote{http://heasarc.gsfc.nasa.gov/docs/suzaku/analysis/abc/abc.html} for details).  
All extractions and analysis were done utilizing HEASOFT v.6.17.


\subsection{XIS Reduction}

The XIS is comprised of 4 CCD's, however XIS2 has been inoperative since 2005 November, when it was likely hit with a 
micrometeorite (see the \textsl{Suzaku} Data Reduction Guide for details).  Two of the remaining CCDs (XIS0 and XIS3) are front-illuminated, 
maximizing the effective area of the detectors in the Fe K bandpass, while the fourth CCD (XIS1) is back-illuminated (BI), 
increasing its effective area in the soft X-ray band ($\lesssim$\,2 keV).  

The XIS events data were taken in 3$\times$3 and 5$\times$5 editing modes, which were cleaned and summed to create image files for each XIS.  
We extracted lightcurves and spectra from a 3 arcmin source region and four 1.5 arcmin background regions.  
The net exposure after screening was 125 ks per XIS.
We used the FTOOLS XISRMFGEN and XISSIMARFGEN to create the response matrix and ancillary response files, respectively.
We then co-added the FI XIS data.
For spectral fitting, XIS-FI and XIS-BI data were ignored above 10 keV where the effective area of the XIS begins to decrease significantly and 
below 0.5 keV due to time-dependent calibration issues of the instrumental O K edge (Ishisaki \etal 2007).

\subsection{HXD Reduction}

The HXD is comprised of two detectors, the PIN diodes (12--70 keV) and the GSO scintillators (50--600 keV).
The PIN is a non-imaging instrument with a 34$\arcmin$ square field of view (FWHM).  
The source was not detected by the GSO, being fainter than the sensitivity limit above 50 keV.
The HXD instrument team provides non-X-ray background (NXB) event files for the PIN using the calibrated GSO data for the particle 
background monitor (``tuned background''), yielding instrument backgrounds with $\lesssim1.5\%$ systematic uncertainty at the 1$\sigma$ 
level (Fukazawa \etal 2009).  We simulated the cosmic X-ray background in \xspec using the form of Boldt (1987), which we combined
with the NXB to created the total PIN background.

PIN spectra were extracted and deadtime-corrected for a net exposure times of 103 ks.
For the purposes of spectral modeling we excluded PIN data below 16 keV due to thermal noise (Kokubun \etal 2007) and above 35 keV where the source is not detected.
The source is only just detected in the 16--35 keV range with the source accounting for 3.7\% of the total counts.
Since this is quite low, we tested how adjusting the background by $\pm$1.5\% affected our results (see below).



\begin{deluxetable*}{lccccccccr}
   \tablecaption{Broadband Pexrav Model Parameters \label{tabpar}}
   \tablecolumns{10}
   \startdata
\hline
\hline\\[-1mm]
 Observed		 				& Photon  		&	Power Law			& 	Fe \ka		& 	Fe \ka				& Fe \ka	& Relative		&  APEC	& APEC	&	$\chi^2$/dof \\[1mm]
 F$_{2-10}$\tablenotemark{A}		& Index 		&	Norm\tablenotemark{B}	& 	Line	Energy	& Line Norm\tablenotemark{C} & $EW$	& Reflection 	& $kT$	& Norm	&		\\[1mm]
							& ($\Gamma$) 	&  						&   	(keV)		&						& (eV)	& ($R$)		& (keV)	& (10\e{-4})			 \\[1mm]
\hline\\
2.24$\,\pm\,$0.02	 			&	1.88$\,\pm\,$0.03	&	6.90$\,\pm\,$0.05	&	6.43$\,\pm\,$0.05	&	1.8$\,\pm\,$0.7	& 77$\,\pm\,$30	&	0.7$\,\pm\,$0.5	&	0.19$\,\pm\,$0.01	&	3.2$\,\pm\,$0.4 	&	435/356	\\[1mm]
\enddata
\tablecomments{Parameters from our best fit model including an unabsorbed power law, a thermal \textsc{apec} component, a Gaussian Fe \ka line (width frozen at 1 eV), and \pexrav Compton-reflection.}
\tablenotetext{A}{Flux is in units of $10^{-12}$~\fluxunits.}
\tablenotetext{B}{Power law norm is in units of $10^{-3}$~counts s\e{-1} keV\e{-1} at 1 keV.}
\tablenotetext{C}{Fe line flux is in units of $10^{-6}$~\feunits.}
\end{deluxetable*}

\begin{figure}
      \includegraphics[width=0.47\textwidth]{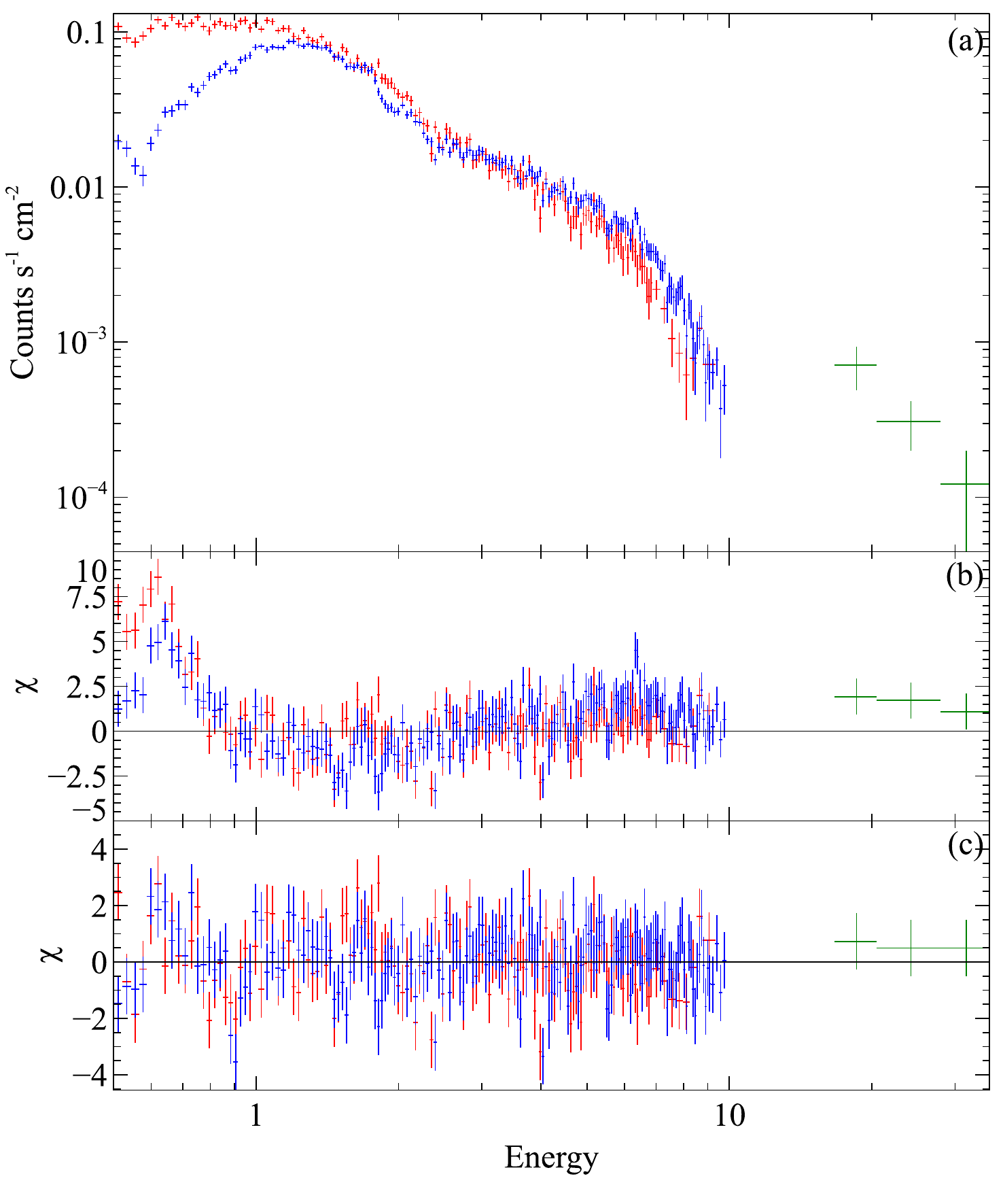}
 \caption{Spectral data from the \suzaku observation of NGC~3660 with residuals to a power law and \pexrav model fit. Panel (a) contains the data with XIS-FI shown in blue, XIS-BI shown in red, and PIN shown in green. Panel (b) shows residuals to a simple power law fit to the 0.5--35 keV range, showing the thermal component below 1 keV, the weak Fe \ka line at 6.4 keV, and the slight excess in the PIN band that may be due to Compton reflection.  Panel (c) shows data minus model residuals to the best fit \pexrav model.}
  \label{figpex}
\end{figure}


\section{Analysis}

We present the XIS combined light curve in Figure \ref{figlc}.  The source shows significant short term variability quite similar to that seen
in the 2009 \xmm observation presented by Bianchi \etal (2012).  We compute a variability amplitude in the 0.5--10 keV range of 0.28$\pm$0.03 with
a fractional variability amplitude, F$_{\rm var}$= ($11\pm1$)\%, typical of unobscured AGN (see, e.g., Nandra et al. 1997).
We can estimate the mass of the supermassive black hole from the excess variance of the light curve 
using the following equation from Ponti \etal (2012): $log(M_{\rm BH,7}) = -1.15^{-1} \times (log(\sigma^2_{\rm RMS, 80\,ks})+1.94)$, 
with expected errors of up to a factor of 5.  Using 80 ks intervals we find an average mass of $M_{\rm BH} \sim 7\times 10^{6}$ \Msun.
This is consistent with estimates from bulge properties, $6.8-21 \times 10^{6}$ \Msun (Bianchi \etal 2012 and references therein).

\subsection{Spectral Analysis}

All spectral fitting was done in \xspec v.12.8.1 (Arnaud 1996) using the solar abundances of Wilms \etal (2000) and cross-sections from Verner \etal (1996). 
We adopt the default cosmological parameter values of $H_{0} = 70$ km s\e{-1} Mpc\e{-1},  $\Omega_{\Lambda} = 0.73$ and $\Omega_{\rm m} = 0.27$.
Uncertainties are listed at the 90\% confidence level ($\Delta \chi^2$ = 2.71 for one interesting parameter).
We included a cross normalization constant for each instrument to account for known cross-calibration uncertainties, 
measuring 0.99 for the XIS-BI and fixed at 1.18 for the PIN relative to the XIS-FI.
and included a Galactic absorption column of 3.5 $\times 10^{20}$ cm\e{-2} in all models (Kalberla \etal 2005).

We started by fitting a simple power law to the 1--35 keV (finding a photon index of $\Gamma \sim$2.0), then expanded our bandpass down to 0.5 keV to examine the broad band residuals.
These residuals (shown in Figure \ref{figpex}b) revealed three distinct features: a soft X-ray excess, a weak Fe \ka line which is only significantly 
detected in the XIS-FI data due to low signal-to-noise in the XIS-BI at higher energies, and a slight excess around 20--30 keV which could be indicative of a Compton reflection hump.

Adding an \textsc{apec} thermal soft X-ray component with Solar abundances improved the fit from \chidof=762/325 for the pure power law to \chidof=433/323.
We also attempted to fit this soft excess with a power law or black body component, but neither fit the data well (\chidof $\sim$ 485/323 with significant residuals below 1 keV).  
We measure a plasma temperature of 0.19 keV for this component with a luminosity of around 10$^{41}$ erg s\e{-1}.
This is quite high compared to the luminosity of the AGN ($L_{2-10} \sim10^{41}$ erg s\e{-1}).  
While it may have an origin in the star forming regions seen previously in the host galaxy (e.g., Bianchi \etal 2012), it is more likely
that the best interpretation is in terms of a 'standard' soft excess for type 1 sources, i.e., warm Comptonization and/or relativistic blurring from the accretion disk, 
however further investigation into the true nature of the soft excess is beyond the scope of this paper.

\begin{figure}
      \includegraphics[width=0.47\textwidth]{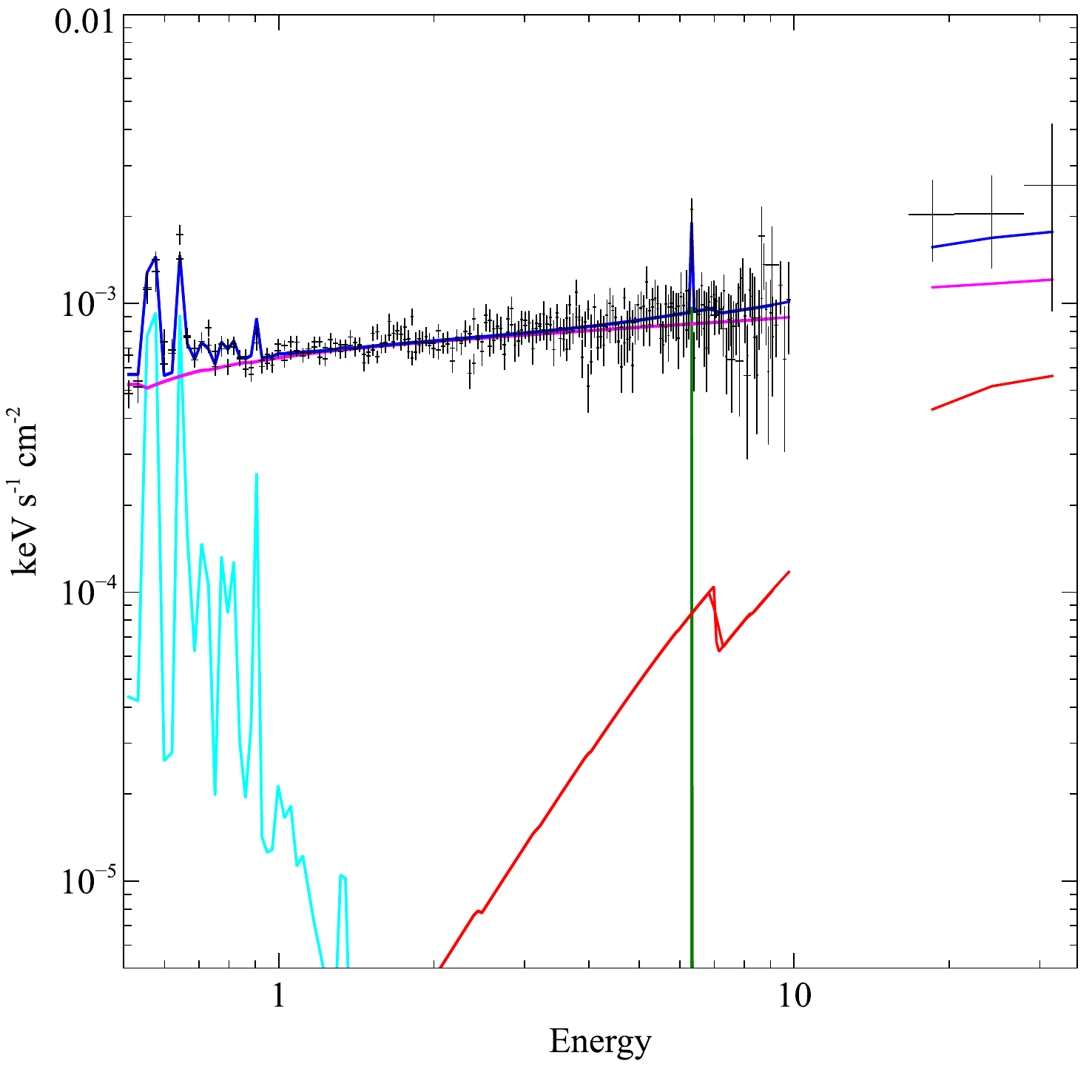}
 \caption{\pexrav model components: power law shown in magenta, \textsl{APEC} shown in cyan, Fe \ka line shown in red, \pexrav shown in red.  Full model shown in blue and unfolded data are shown in black.}
  \label{figeuf}
\end{figure}

Including a narrow Gaussian Fe \ka line improves the fit further to \chidof=409/321 with a 
well measured line energy (6.43 keV, consistent with an origin in neutral material) but a relatively weak equivalent width, $EW$=77$\,\pm\,$30 eV.
We next added a Compton reflection hump using the \pexrav model (Magdziarz \& Zdziarski 1995).
This improved the fit marginally (\chidof=404/320) with a slight visual improvement to the PIN residuals.
We chose to freeze the inclination angle at 65\degr as would be expected for a typical Seyfert 2, though given the lack of absorption along the line of sight
and lack of intrinsic broad line emission the inclination angle may be much lower. 
Parameters for our best fit model are presented in Table 1, and the data minus model residuals are shown in Figure \ref{figpex}c.
Model components are shown in Figure \ref{figeuf}.
We also achieved a good fit with the \pexmon model (\chidof=407/322) which includes the Fe \ka line self-consistently modeled as part of the reflection spectrum.

We also attempted to fit the \mytorus model (Murphy \& Yaqoob 2009) to the XIS$+$PIN spectrum, including the thermal \textsc{apec} component.
We froze the inclination angle to 90\degr\ and tied \nh between the absorbed continuum, Compton scattered component, and Fe emission complex.
We tied the normalizations of the Fe line and the scattered component together, but left them independent from the normalization of the absorbed continuum. 
The excess in the PIN band is not very large and consequently the \mytorus parameters are not well constrained,
particularly the column density, so we constrained it to be Compton-thick ($>10^{24}$ \colunits).  
This fit was very poor (\chidof$\sim$3000/323).

Adding a ``leaked'' power law to the above \mytorus model with $\Gamma$ tied and normalization limited to $\lesssim$10\% of the continuum flux as would be expected for 
scattering in a heavily obscured toroidal geometry led to an improved but still very poor fit (\chidof$\sim$1000/322).
Allowing for a free normalization of the leaked power law we find a reasonable fit (\chidof=416/322) with a normalization of 
($7.0\pm0.1$)\ten{-4} counts cm\e{-2} s\e{1} at 1 keV compared to a normalization of the absorbed component of ($5\pm3$)\ten{-4} counts cm\e{-2} s\e{1} at 1 keV.
We measure an intrinsic $F_{2-10}=4\ten{-12}$ \fluxunits.
Freezing the \mytorus inclination to 0\degr\ for a face-on unobscured geometry gives a good fit (\chidof=409/323) 
with an equatorial column density of the torus of \nh$\,\sim\,2.3$\err{4.9}{1.8} \ten{24} \colunits.

We tested the influence of the PIN background estimation on the \mytorus modeling by varying the background flux by $\pm$1.5\%.
This had little effect on the upper limit of the absorbed component, instead increasing/decreasing the scattered component (i.e., the reflection hump).
Tying the scattered normalization to the heavily absorbed power law normalization leads to poor residuals in the PIN band due to mismatching between the PIN and XIS.

\section{Discussion}

NGC~3660 is optically classified as a type 2 Seyfert with no evidence for changing state behavior  
and simultaneous optical and \xmm observations that rule out the possibility of a mismatch (Bianchi \etal 2012).
X-ray observations have typically revealed a Seyfert 1-like unabsorbed X-ray spectrum.
However, a \sax observation in the hard X-ray band showed a strong excess above 10 keV that could be indicative of a 
nucleus embedded in Compton-thick material sufficient to completely obscure the BLR as well as the majority of the X-ray flux.

Our \suzaku results do not show the same hard X-ray excess as the \sax data, with an upper limit to the 20--100 keV flux 
of $\sim$8\ten{-12} \fluxunits ($\sim$1.2\ten{-11} \fluxunits when PIN background uncertainties are taken into account), 
an order of magnitude lower than the \sax measurement.  The slight excess above the power law that we do see is 
consistent with reflection from the accretion disk or torus typical of an unobscured Seyfert.  Attempting to fit the spectrum with a heavily obscured torus model gave an 
intrinsic flux of such an embedded nucleus nearly equal to the flux of the observed $<10$ keV X-ray power law, that is to say a covering fraction of $\sim$\,50\%.
Both the heavily obscured and unobscured models give intrinsic luminosities consistent with the measured bolometric luminosity of $L_{\rm bol} \sim 10^{43}$ erg s\e{-1} 
($L_{2-10}=1-2 \times 10^{41}$ erg s\e{-1} corresponds to a bolometric luminosity of $L_{\rm Bol} \sim 3-6 \times 10^{42}$ erg s\e{-1}, see, e.g., Vasudevan \etal 2010).

We also measured the Fe \ka line emission to be far weaker ($EW=77\,\pm\,30$ eV) than would be expected in a Compton-thick, reflection-dominated scenario.
The flux of this line is consistent with the Fe \ka line measurement from the \xmm spectrum of (1.8$\pm$1.5)\ten{-6} \fluxunits (Bianchi \etal 2012).
The neutral nature of this line indicates that it likely arises in distant, cold material.

The source has not been detected by the \bat hard X-ray survey, nor was it detected by \xte in the 3--250 keV band,
indicating a typical flux much lower than that measured by \sax, $F_{14-195} \lesssim 1 \times 10^{-11}$ \fluxunits, assuming our best fit model.
This is consistent with our measured flux in the hard band which would correspond to $F_{14-195} \sim 6 \times 10^{-12}$ \fluxunits).  
A possible explanation for the disagreement with \sax PDS measurements is a hard X-ray transient in the \sax field of view, though this is difficult to confirm.

Shi \etal (2010) had ruled out NGC 3660 as a true Seyfert 2 candidate, noting that the broad ($\sim2000$ km s\e{-1}) H$\alpha$ component
may be weak ($L_{\rm H\alpha} \sim 10^{40}$ erg s\e{-1}) but that confusion from HII regions was responsible for the mistyping.   
Bianchi \etal (2012) obtained several recent optical spectra and detected a tentative broad component with 
$v \sim 3000$ km s\e{-1} and $L_{\rm H\alpha} = 6\times10^{39}$ erg s\e{-1}, two orders of magnitude lower than the 
expected value of $3\times10^{41}$ erg s\e{-1} based on the bolometric luminosity
(note that typical dispersions in this relation are $\sim$0.5 dex; see, e.g., Stern \& Laor 2012).

Trump \etal (2011) theorized that a BLR cannot form below a certain accretion rate.
NGC 3660 has a low Eddingtion ratio, $L_{\rm Bol}/L_{\rm Edd}\,\sim\,0.4-2$\ten{-2} (Bianchi \etal 2012), 
consistent with predictions by Trump \etal (2011), though much higher than the Eddington ratio of, for example, NGC 3147.
This could be the reason for the weakness of the BLR measured in this source.

\begin{acknowledgments}

This research has made use of data obtained from the \textsl{Suzaku} satellite, a collaborative 
mission between the space agencies of Japan (JAXA) and the USA (NASA).
This work has made use of HEASARC online services, supported by NASA/GSFC, and the NASA/IPAC Extragalactic Database, 
operated by JPL/California Institute of Technology under contract with NASA.
This work was supported under NASA Contract No. NNG08FD60C and sub-contract No. 44A-1092750.

\end{acknowledgments}

{\it Facilities:} \facility{Suzaku}



\end{document}